\documentclass[a4paper,12pt]{article}
\usepackage{graphicx}
\usepackage{amssymb}

\begin{document}
\baselineskip=15.5pt
\pagestyle{plain}
\setcounter{page}{1} 

\setlength\arraycolsep{2pt}

\begin{titlepage}


\rightline{\small{DESY 07-068}}

\begin{center}

\vskip 1.7 cm

{\LARGE {\bf On Newton's law in supersymmetric braneworld models}}

\vskip 1.5cm
{\large 
Gonzalo A. Palma
}
\vskip 1.2cm

Deutsches Elektronen-Synchrotron DESY, Theory Group, 
Notkestrasse 85, D-22603 Hamburg, Germany

\vskip 0.5cm

\vspace{1cm}

{\bf Abstract}

\end{center}

We study the propagation of gravitons within 5-D supersymmetric braneworld models
with a bulk scalar field. The setup considered here consists of a 5-D bulk 
spacetime bounded by two 4-D branes 
localized at the fixed points of an $S^{1}/\mathbb{Z}_{2}$ orbifold. There 
is a scalar field $\phi$ in the bulk which, provided a superpotential $W(\phi)$, 
determines the warped geometry of the 5-D spacetime. 
This type of scenario is common in string theory, where the bulk scalar field $\phi$ 
is related to the volume of small compact extra dimensions. 
We show that, after the moduli are stabilized by supersymmetry breaking terms localized 
on the branes, the only relevant degrees of freedom in the bulk consist 
of a 5-D massive spectrum of gravitons. Then we analyze the gravitational 
interaction between massive bodies localized at the positive tension brane
mediated by these bulk gravitons. It is shown that the Newtonian potential 
describing this interaction picks up a non-trivial contribution at short distances
that depends on the shape of the superpotential $W(\phi)$. We compute
this contribution for dilatonic braneworld scenarios
$W(\phi) \propto e^{\alpha \phi}$ (where $\alpha$ is a constant) and discuss
the particular case of 5-D Heterotic M-theory: It is argued that a specific 
footprint at micron scales could be observable in the near future.

\noindent

\end{titlepage}

\newpage

\section{Introduction}

Recent tests of gravity at short distances \cite{Kapner:2006si, Hoyle:2000cv, 
Chiaverini:2002cb, Adelberger:2003zx} 
have confirmed that Newton's inverse-square law holds down to a length 
scale $56 \mu$m. This has substantially improved previous 
constraints on exotic interactions mediated by the exchange of 
massive scalars or vectors between neutral atoms \cite{Adelberger:2006dh}, where a 
Yukawa type contribution to the Newtonian potential is generally expected.
It has also lowered the scale at which large extra dimensions 
\cite{Antoniadis, Arkani-Hamed:1998rs, Antoniadis:1998ig}
and braneworld models \cite{Akama:1982jy, Randall:1999ee, Randall:1999vf} 
may show up by affecting the propagation of 
gravitons in the presence of a large --or infinite-- extra dimensional 
volume.


Indeed, in theories where matter fields confine to a 4-D brane 
and gravity is the only massless 
field able to propagate along the extra dimensional volume 
\cite{Langlois:2002bb, Maartens:2003tw, Brax:2004xh}, one 
generally expects short distance corrections to the usual 4-D Newtonian potential. 
The shape and distance at which these corrections become relevant generally depend 
on the geometry and size of the extra dimensional volume, thus 
allowing for distinctive signals dependent of the particular content of 
the theory. In the single-brane Randall-Sundrum scenario \cite{Randall:1999vf}, 
for instance, where a 4-D brane of constant tension $\propto k$ 
is immersed in an infinitely large five-dimensional
AdS volume, a zero mode graviton $g_{\mu \nu}$ localizes about the brane. 
This zero mode is exponentially suppressed away from the brane 
with a warp factor $\propto e^{-kz}$, where $z$ is the 
distance from the brane along the fifth extra-dimensional direction. 
The Newtonian potential
describing the gravitational interaction between two bodies 
of masses $m_{1}$ and $m_{2}$ localized at the brane, and separated 
by a distance $r$, is then found to be \cite{Randall:1999vf, Garriga:1999yh,
Kiritsis:2002ca, Ghoroku:2003bs, Callin:2004py}
\begin{eqnarray} \label{c1: V(r)}
V(r) = - G_{\mathrm{N}} \frac{m_{1} m_{2}}{r} \Big( 1 + \frac{2}{3 k^{2}
r^{2}} \Big), \label{V-RS}
\end{eqnarray}
where $G_{\mathrm{N}}$ is Newton's constant. The correction $2/3 k^2 r^2$ springs out directly 
from the way in which bulk gravitons propagate in an AdS five-dimensional spacetime.
A correction like this provides an important signature for the low energy phenomenology 
of braneworld models with warped extra-dimensions; if the tension
$k$ is small enough as compared to the Planck mass 
$M_{\mathrm{Pl}} = (8 \pi G_{\mathrm{N}})^{-1/2}$, 
then it would be possible to distinguish this type of scenario from other 
extra-dimensional models in up-coming short distance tests of gravity
(present tests give the robust constraint $1/k < 11 \mu$m). 

It is therefore sensible to ask how other braneworld scenarios may differ from 
the Randall-Sundrum case at short distances, especially within the context of 
more realistic models.
The purpose of this paper is to shed light towards this direction.
Here we study the propagation of gravitons 
within 5-D braneworld models where the geometry of the extra-dimensional 
space differs from the usual AdS profile. We will show that the gravitational
interaction at short distances is sensitive to the geometry of the 
extra-dimensional bulk in such a way that the Newtonian potential 
picks up a non-trivial correction at scales comparable to the tension of 
the brane. As we shall see, this correction may differ dramatically from 
the one depicted in Eq. (\ref{V-RS}). We refer to 
\cite{Nojiri:2000yr, Ito:2001nc, Callin:2004bm, Arnowitt:2004vq, Callin:2005wi, deRham:2006hs, 
Bronnikov:2006jy, Buisseret:2007qd} 
for other works on short distance modifications to general relativity 
within the braneworld paradigm.

\subsection{General idea}

We will look into a fairly general class of supersymmetric braneworld 
scenarios with a bulk scalar field $\phi$. 
The model considered here consists of a 5-D bulk 
spacetime bounded by two 4-D branes localized at the fixed points 
of an $S^{1}/\mathbb{Z}_{2}$ orbifold. The tensions of the branes are 
proportional to the superpotential $W(\phi)$ of the theory, allowing for 
BPS configurations in which half of the bulk supersymmetry is broken 
on the branes \cite{Bergshoeff:2000zn, Brax:2000xk}.
These types of models are well motivated from string theory, 
particularly within the heterotic M-theory approach \cite{Horava:1995qa, Horava:1996ma} 
where, in the 5-D effective low energy theory, 
the scalar field $\phi$ is related to the size of the volume of small extra-dimensions 
compactified on a Calabi-Yau 3-fold \cite{Lukas:1998yy, Lukas:1998tt}. 
To gain insight into the gravitational phenomenology of this model,
we shall only consider the bosonic sector of the theory.

One crucial property for us coming from this class of models 
is that the warping of the extra-dimensional volume 
depends on the form of the superpotential $W(\phi)$. To be more 
precise, given a metric $g_{\mu \nu} = \omega^{2}(z) \eta_{\mu \nu}$, where 
$\eta_{\mu \nu}$ is the usual Minkowski metric and $z$ is the coordinate 
parameterizing the proper distance along the extra-dimension,
then bulk fields are related to $W(\phi)$ by means of the following
first order differential equations
\begin{eqnarray}
\frac{\omega'}{\omega} = - \frac{1}{4} W \qquad \mathrm{and} \qquad 
\phi' = \frac{\partial W}{\partial \phi}, \label{BPScond1}
\end{eqnarray}
where $' \equiv \partial_{z}$. In the particular case of Randall-Sundrum branes, 
the superpotential is simply a constant $W = 4 k$, implying an AdS5 
spacetime with $\omega = e^{-kz}$. For more general superpotentials, 
however, the warping of the extra-dimension may have a richer 
structure and even contain singularities \cite{Maldacena:2000mw}.
For example, in the case of dilatonic braneworlds, 
where $W(\phi) = \Lambda \, e^{\alpha \phi}$, one encounters a singularity 
$\omega=0$ at a distance $z = 1/ \alpha^{2} W(\phi_{1})$ 
from the positive tension brane, where $\phi_{1}$ is 
the value that the scalar field acquires on the brane. 
In this way, while in the Randall-Sundrum model there is a
positive-tension brane in an infinite volume 
(with $\omega=0$ at infinity), in more general
cases one may have a single positive-tension brane immersed in a 
finite bulk-volume at a certain distance from the $\omega=0$ singularity.

The basic idea of this paper is to compute the effect of such geometries on 
the gravitational interaction between massive bodies localized on 
the same brane. For simplicity, we shall consider single-brane configurations
in which the visible brane corresponds to the positive tension 
brane, whereas the negative tension brane localizes at the 
bulk singularity.

\subsection{The moduli problem}

Models with extra-dimensions generically predict 
the existence of massless degrees of freedom, the moduli, at the 4-D effective theory 
level \cite{Garriga:1999yh, Chiba:2000rr}. 
These moduli appear coupled to the matter sector with the 
same strength as gravity, leading to significant long range modifications 
to general relativity, well constrained by 
both Solar system \cite{Will:2001mx, Cassini, Lunar} and binary pulsar 
tests \cite{Damour:1996ke, Damour:1998jk, 
Weisberg:2002nv, Esposito-Farese:2004cc}. In the Randall-Sundrum model, 
the radion moduli vanishes as the negative tension brane 
disappears at infinity --the single brane limit-- leaving gravity
as the only relevant long range interaction of the model.
However, in more general braneworld scenarios --as the one we consider here-- 
one typically expects other massless degrees of freedom, even in the single brane limit.
For instance, current Solar system tests imply 
a constraint on the following dimensionless parameter 
$\alpha \equiv W^{-1} \partial_{\phi} W$ 
\begin{eqnarray}
\alpha^{2} < 1.5 \times 10^{-6}, \label{constr}
\end{eqnarray}
where $\alpha$ is evaluated on the brane in which 
tests are performed \cite{Palma:2005wm}. The Randall-Sundrum model corresponds to the
trivial case $\alpha^{2}=0$, thus passing the test with flying colors,
nevertheless, in more realistic 
models one has $\alpha^{2} \simeq \mathcal{O}(1)$. 
As we shall learn later in more detail, in order to have significant 
effects at short distances --say, at micron scales-- different 
from the Randall-Sundrum case, it is necessary to be in the 
range $\alpha^{2} \simeq \mathcal{O}(1)$. This strongly contrasts with 
the bound of Eq. (\ref{constr}).

One way out of this problem consists in taking into account 
a stabilization mechanism for the moduli \cite{Goldberger:1999uk, Tanaka:2000er}, 
in this case, 
for the bulk scalar field $\phi$. If $\phi$ becomes massive on the branes, 
then the only relevant long range interaction in the bulk would consists 
of the gravitational field. To this extent, we consider supersymmetry 
breaking potentials localized at the orbifold fixed points. We will show that, 
provided certain simple conditions on these potentials, it is possible 
to stabilize the bulk scalar field $\phi$ without spoiling the vacuum geometry 
of the extra-dimensional space dictated by Eq. (\ref{BPScond1}), and therefore 
retaining all the interesting features coming from the bulk-curvature.

\subsection{Plan of the paper}

This work is organized in the following way:
We start in Section \ref{superbranes} by introducing braneworld 
models with a bulk scalar field. There, we deduce the equations of motion 
of the system and present the zero mode background solution --a BPS vacuum state-- 
and its effective theory, a bi-scalar-tensor theory of gravity. 
The scalar degrees of freedom of this theory consist of the boundary values of $\phi$ at 
both branes.\footnote{Alternatively, one may define the two scalar 
degrees of freedom as the distance between the branes (the radion) 
plus just one boundary value of $\phi$ at a given brane.}
Then in Section \ref{linear} we study the linear perturbations
of the fields around the zeroth-order solution presented in Section 
\ref{superbranes}. There, we also consider the 
problem of stabilizing the moduli. We show that, once the zeroth-order 
moduli are stabilized, the only relevant degrees of freedom at low energies 
in the bulk consist of a massive tower of 5-D gravitons. 
In Section \ref{Newtonian pot} we analyze the effects of these massive 
states on the gravitational interaction between massive bodies localized at the 
same brane. There, we compute the short distance modifications to 
Newton's inverse-square law parameterized by a function $f(r)$ 
appearing in the form
\begin{eqnarray}
V(r) = - G_{\mathrm{N}} \frac{m_{1} m_{2}}{r} \left[ 1 
+ f(r) \right].
\end{eqnarray}
As we shall see, the shape of the function $f(r)$ depends heavily on the shape 
of the superpotential $W(\phi)$. To put things into context, we further 
compute the corrections arising in the particular case of dilatonic braneworlds 
$W(\phi) \propto e^{\alpha \phi}$, where $\alpha$ is a constant. 
There, we also discuss the particular case of 5-D Heterotic M-Theory and
the prospects of observing these corrections 
in the near future. Finally, in Section \ref{Conclusions} we 
provide some concluding remarks.

\section{Braneworld models with a bulk scalar field} \label{superbranes}

\begin{figure}[h]
\begin{center}
\includegraphics[width=0.6\textwidth]{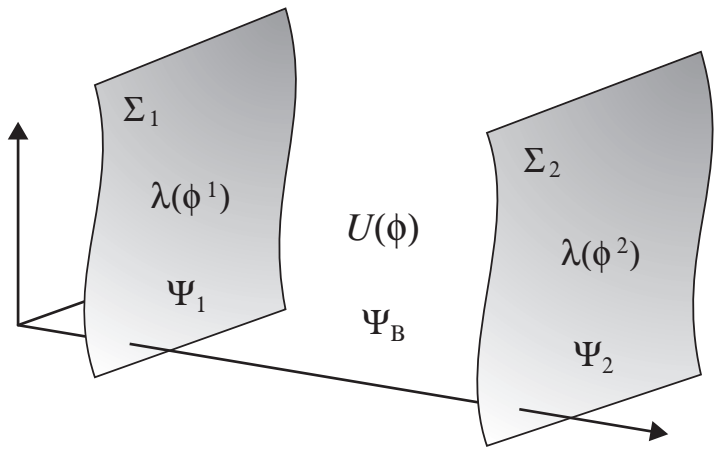}
\caption[Basic configuration]{\footnotesize Schematic representation
of the 5-D brane configuration. In the bulk there is a scalar field
$\phi$ with a bulk potential $U(\phi)$.
Additionally, the bulk space is bounded by branes $\Sigma_{1}$ and
$\Sigma_{2}$ located at the orbifold fixed points. The branes are
characterized by tensions $\lambda_{1}$ and $\lambda_{2}$, and
contain matter fields $\Psi_{1}$ and $\Psi_{2}$ respectively. }
\label{fig: chap2-2}
\end{center}
\end{figure}

Let us consider a 5-D spacetime $M$ with topology $M = \mathbb{R}^{4}
\times S^{1}/ \mathbb{Z}_{2}$, where $\mathbb{R}^{4}$ is a fixed 4-D
Lorentzian manifold without boundaries and $S^{1}/\mathbb{Z}_{2}$ is
the orbifold constructed from the one-dimensional circle with points
identified through a $\mathbb{Z}_{2}$-symmetry. $M$ is bounded 
by two 3-branes located at the fixed points of
$S^{1}/\mathbb{Z}_{2}$. We denote the brane hyper-surfaces by
$\Sigma_{1}$ and $\Sigma_{2}$ respectively, and call the space $M$
bounded by the branes, the bulk space. In this model there is a bulk
scalar field $\phi$ with a bulk potential $U(\phi)$ and boundary
values $\phi^{1}$ and $\phi^{2}$ at the branes. Additionally, the
branes have tensions $\lambda_{1}$ and $\lambda_{2}$ which are given
functions of $\phi^{1}$ and $\phi^{2}$, respectively 
(see Fig. \ref{fig: chap2-2}).
The total action of the system is given by
\begin{eqnarray}
S_{\mathrm{tot}} = S_{\mathrm{bulk}} + S_{\mathrm{brane}}, \label{c2: tot-act}
\end{eqnarray}
where $S_{\mathrm{bulk}}$ is the term describing the gravitational physics at the bulk
(including the bulk scalar field)
\begin{eqnarray}
S_{\mathrm{bulk}} = \frac{M_{5}^{3}}{2} \int_{M} \! R^{(5)} 
- \frac{3 M_{5}^{3}}{8} \int_{M} \! 
\left[  (\partial \phi)^{2} + U(\phi) \right] + S_{\mathrm{GH}}. 
\label{c2: S-g}
\end{eqnarray}
Here the integral symbol $\int$ is the short notation for $\int d^{5}X \sqrt{- g_{5}}$, 
where $X^{A}$, with $A= 1,\cdots,5$, is the coordinate system covering $M$ and 
$g_{5}$ is the determinant of the 5-D metric $g_{A B}$ of signature $(-++++)$. 
$M_{5}$ is the 5-D fundamental mass scale and $R^{(5)}$
is the 5-D Ricci scalar. Observe that in the present notation
the bulk scalar field $\phi$ is dimensionless.
The third term of Eq. (\ref{c2: S-g}) corresponds to the 
Gibbons-Hawking boundary term 
$S_{\mathrm{GH}} \propto \int_{\Sigma_{1}} K - \int_{\Sigma_{2}} K$, added
to make the bulk gravitational physics regular near the fixed points.
The action $S_{\mathrm{brane}}$ appearing in 
Eq. (\ref{c2: tot-act}) stands for the fields at the fixed points. 
It is given by
\begin{eqnarray} 
S_{\mathrm{brane}} =  S_{\mathrm{matter}} 
- \frac{3 M_{5}^{3}}{2} \, \int_{\Sigma_{1}}
\! \lambda_{1} (\phi^{1}) - \frac{3 M_{5}^{3}}{2} \,
\int_{\Sigma_{2}} \! \lambda_{2} (\phi^{2}) , \label{c2: brane-tensions}
\end{eqnarray}
where $\lambda_{1}(\phi^{1})$ and $\lambda_{2}(\phi^{2})$ are the
brane tensions and $S_{\mathrm{matter}}$ the
action describing the matter content of the branes, which we write
\begin{eqnarray}
S_{\mathrm{matter}} = S_{1}[\Psi_{1}, g_{\mu \nu}^{(1)}] +
S_{2}[\Psi_{2},g_{\mu \nu}^{(2)}], \label{c2: brane-matter}
\end{eqnarray}
where $\Psi_{1}$ and $\Psi_{2}$ denote the matter fields at each
brane, and $g_{\mu \nu}^{(1)}$ and $g_{\mu \nu}^{(2)}$ are 
the induced metrics at $\Sigma_{1}$ and $\Sigma_{2}$ respectively. 
In what follows we summarize some important properties of this 
system.

\subsection{5-D supergravity} \label{supergravity}

As already mentioned, we focus our interest on a class of models embedded in 
supergravity, where the bulk potential $U(\phi)$ and the brane 
tensions $\lambda_{1}(\phi^{1})$ and $\lambda_{2}(\phi^{2})$ satisfy
a special relation so as to preserve half of the
local supersymmetry near the branes \cite{Bergshoeff:2000zn}. The relation turns out to be
\begin{eqnarray}
U =  (\partial_{\phi} W)^{2} - W^{2}, \qquad
\lambda_{1} = W(\phi^{1}), \qquad \mathrm{and} \qquad
\lambda_{2} = - W(\phi^{2}), \label{BPS cond}
\end{eqnarray}
where $W = W(\phi)$ is the superpotential of the system. Observe that 
the tensions $\lambda_{1}$ and $\lambda_{2}$ depend on $W$ with opposite signs.

Several aspects of this class of models have been thoroughly
investigated over the last few years, among them: Braneworld inflation 
\cite{Lukas:1999yn, Brax:2002nt, Ringeval:2005yn}, their low energy dynamics 
\cite{Mukohyama:2001ks, Brax:2001fh, Kobayashi:2002pw, Palma:2004fh, 
Palma:2004et, Brax:2004bw, Lehners:2006ir},
brane collisions \cite{Webster:2004sp, Lehners:2006pu}, 
and various phenomenological aspects \cite{Palma:2003rs, Brax:2004ym, Zanzi:2006xr}.
This class of model is attractive for several reasons: On the one 
hand, they offer a natural extension
to the much studied Randall-Sundrum model, 
where a fine tuning condition between
the bulk cosmological constant and the tensions allows a null effective 
cosmological constant on the brane. This is also the case here  
\cite{DeWolfe:1999cp, Csaki:2000wz, Binetruy:2000wn, Flanagan:2001dy} where
condition (\ref{BPS cond}) implies a zero effective dark energy term on the brane. 
In fact, the case $W=$ constant corresponds to the particular case of
Randall-Sundrum branes. On the other hand, this is the generic class of models one 
would expect from superstring theories, where the size of the volume of 
compactified extra-dimensions is modeled as a scalar field.
For example, in low energy Heterotic M-theory it is found, after compactifying 
6 of the 10 spatial dimensions on a Calabi-Yau 3-fold \cite{Lukas:1998tt}, 
a superpotential of the form $W (\phi) \propto e^{\alpha \phi}$, 
with $\alpha^{2} = 3/2$.

Since in the real world supersymmetry is expected to be broken, 
it is convenient to consider small deviations 
from the configuration of Eq. (\ref{BPS cond}). 
We do this by introducing supersymmetry breaking 
potentials $v_{1}(\phi^{1})$ and $v_{2}(\phi^{2})$ at the branes in
the following way
\begin{eqnarray}
U =  (\partial_{\phi} W)^{2} - W^{2}, \qquad
\lambda_{1} = W + v_{1}, \qquad \mathrm{and} \qquad
\lambda_{2} = - W - v_{2},
\end{eqnarray}
with $|v_{1}| \ll |W(\phi^{1})|$ and $|v_{2}| \ll |W(\phi^{2})|$. 
Potentials $v_{1}(\phi^{1})$ and $v_{2}(\phi^{2})$
parameterize deviations from the BPS condition (\ref{BPS cond}). 
The precise mechanism by which they are generated is out of the scope of
the present work. We refer to \cite{Brax:2003dh, Brax:2004zs} 
for discussions on this issue.

\subsection{4-D covariant formulation} \label{c2: sec-cov}

Given the topology $M=\mathbb{R}^{4} \times S^{1}/ \mathbb{Z}_{2}$, 
it is convenient to decompose the coordinate
system $X^{A}$ into $(x^{\mu} , z)$, where $x^{\mu}$ with $\mu=1,\cdots,4$
covers the $\mathbb{R}^{4}$ foliations and surfaces $\Sigma_{1}$ and $\Sigma_{2}$,
and where $z$ covers the $S^{1}/\mathbb{Z}_{2}$ orbifold and 
parameterizes the 4-D foliations. With this decomposition, it is customary to write
the metric line element $ds^{2} = g_{A B} dX^{A} dX^{B}$ as  
\begin{eqnarray} \label{c2: ADM-metric}
ds^{2} = N^{2} dz^{2} +  g_{\mu \nu} (dx^{\mu} + N^{\mu} dz)
(dx^{\nu} + N^{\nu} dz) .
\end{eqnarray}
Here, $N$ and $N^{\mu}$ are the lapse and shift functions for the
extra dimensional coordinate $z$, and $g_{\mu \nu}$ is the induced
metric on the 4-D foliations with a $(-+++)$ signature. At the
boundaries we have $g_{\mu \nu}^{(1)} = g_{\mu \nu}(z_{1})$ and
$g_{\mu \nu}^{(2)} = g_{\mu \nu}(z_{2})$. It is possible to show
that the unit-normal vector $n^{A}$ to the foliations has components
\begin{eqnarray}
n^{A} = (-N^{\mu}/N,1/N), \qquad  n_{A} = (0,N).
\end{eqnarray}
Additionally, it is useful to define the extrinsic curvature $K_{\mu \nu}$
of the 4-D foliations as 
\begin{eqnarray}
K_{\mu \nu} = \frac{1}{2N} \left[ g_{\mu \nu}' - \nabla_{\mu}
N_{\nu} - \nabla_{\nu} N_{\mu} \right],
\end{eqnarray}
where  $' = \partial_{z}$ and covariant derivatives $\nabla_{\mu}$ are
constructed from the induced metric $g_{\mu \nu}$ in the standard
way. Another way of writing the extrinsic curvature is 
$2 N K_{\mu \nu} = \pounds_{N n} g_{\mu \nu}$, where 
$\pounds_{N n}$ is the Lie derivative along the vector field
$N n^{A}$.

The present notation allows us to reexpress $S_{\mathrm{bulk}}$ of Eq. (\ref{c2: S-g}) in the
following way
\begin{eqnarray}
S_{\mathrm{bulk}} &=& \frac{M_{5}^{3}}{2} \int_{S^{1}\!/\mathbb{Z}_{2}}
\!\!\!\!\!\!\! dz \! \int \!\! d^{\, 4} x \sqrt{-g} \, N \bigg( R - [
K_{\mu \nu} K^{\mu \nu} - K^{2} ]  - \frac{3}{4} \big[
(\phi'/N )^{2} \nonumber\\ 
&& +  (g^{\mu \nu} + N^{\mu} N^{\nu}/N^{2}) \partial_{\mu} \phi \partial_{\nu}  
\phi  - 2 N^{-2} N^{\mu} \, \partial_{\mu} \phi \, \partial_{z} \phi  + U \big] \bigg),  
\label{c2: cov-G}
\end{eqnarray}
where $R$ is the four-dimensional Ricci scalar constructed from
$g_{\mu \nu}$ and $K=g^{\mu \nu} K_{\mu \nu}$ is the trace of the 
extrinsic curvature. Observe that the Gibbons-Hawking boundary term
$S_{\mathrm{GH}}$, which appeared originally in $S_{\mathrm{bulk}}$,
has been absorbed by the use of metric (\ref{c2: ADM-metric}). 
Let us clarify here that the integration in Eq. (\ref{c2: cov-G})
along the fifth-dimension is performed on
the entire circle $S^{1}$, instead of just half of it. We should
keep in mind, however, that degrees of freedom living in different
halves of the circle are identified through the
$\mathbb{Z}_{2}$-symmetry.

\subsection{Dynamics and boundary conditions}

In this section we deduce the equations of motion governing the
dynamics of the fields living in the bulk and the branes. These
equations are obtained by varying the total action of the system
$S_{\mathrm{tot}}$ with respect to the bulk gravitational fields
$N$, $N^{\mu}$, $g_{\mu \nu}$ and $\phi$, taking special care on the variation of
the boundary terms. The brane tensions
$\lambda_{1}$ and $\lambda_{2}$ and matter fields $\Psi_{1}$ and
$\Psi_{2}$ play a decisive role in determining boundary conditions on the
bulk gravitational fields at $\Sigma_{1}$ and $\Sigma_{2}$,
respectively. The variation of $S_{\mathrm{tot}}$ with respect to $N$, $N^{\mu}$ and
$g_{\mu \nu}$ respectively, gives 
\begin{eqnarray}
R + \left[ K_{\mu \nu} K^{\mu \nu} - K^{2} \right] = - \frac{3}{4} \bigg[
\frac{1}{N^{2}} (\phi')^{2} - g^{\mu \nu} \partial_{\mu} \phi \partial_{\nu}  \phi
- U \bigg], \label{c2: Grav-eq1}
\\
\nabla_{\mu} \left[ K^{\mu}_{\nu} - K \delta^{\mu}_{\nu} \right] = \frac{3}{4}
\frac{1}{N} \phi' \partial_{\mu} \phi , \label{c2: Grav-eq2}
\\
G_{\mu \nu} = \frac{1}{N} ( \nabla_{\mu} \nabla_{\nu} N - g_{\mu \nu} \Box N ) 
- \frac{1}{2} g_{\mu \nu} \left[ K_{\rho \sigma} K^{\rho \sigma} + K^{2} \right] 
+\frac{1}{N} (K_{\mu \nu} - g_{\mu \nu} K )' \nonumber\\ + 3 K K_{\mu \nu} 
- 2 K_{\mu \alpha} K^{\alpha}_{\,\, \nu}
- \frac{3}{8} g_{\mu \nu} \left[  (\phi'/N)^{2} 
+ g^{\mu \nu} \partial_{\mu} \phi \partial_{\nu}  \phi +  U \right] 
+ \frac{3}{4} \partial_{\mu} \phi \partial_{\nu} \phi. \label{c2:
Grav-eq3}
\end{eqnarray}
To write these equations we have adopted Gaussian normal coordinates, which
correspond to the gauge choice $N^{\mu} = 0$ (with this gauge 
one has $\pounds_{N n} = \partial_{z}$).
Here $G_{\mu \nu} = R_{\mu \nu} - \frac{1}{2} g_{\mu \nu} R$ is the Einstein tensor constructed
out from the induced metric $g_{\mu \nu}$. The bulk scalar field equation of motion can
be deduced either from the previous set of equations 
(by exploiting energy momentum conservation), or just by varying 
the action $S_{\mathrm{tot}}$ with respect to $\phi$. One obtains
\begin{eqnarray}
\left[ \nabla^{\mu} \left( N
\partial_{\mu} \phi \right) + \left( \phi' / N \right)' +
K \phi' \right]  = \frac{N}{2} \frac{\partial U}{\partial \phi} . \label{c2: scalar}
\end{eqnarray}
Near the branes the variation of the action leads to a set
of boundary conditions known as the Israel matching
conditions \cite{Israel}. In the present model, they are given by
\begin{eqnarray}
 K_{\mu \nu} - g_{\mu \nu} K &=&  \frac{3}{4} \lambda_{1} \, g_{\mu \nu}
  - \frac{1}{2} M_{5}^{-3} T_{\mu \nu}^{1}, \label{c2: Israel-matching2-s1} \\
\phi' &=&  N \, \partial_{\phi} \lambda_{1}, \label{c2:
BPS-matching2-s1}
\end{eqnarray}
at the first brane $\Sigma_{1}$, and
\begin{eqnarray}
 K_{\mu \nu} - g_{\mu \nu} K &=&  - \frac{3}{4} \lambda_{2} \, g_{\mu \nu}
  + \frac{1}{2} M_{5}^{-3} T_{\mu \nu}^{2}, \label{c2: Israel-matching2-s2} \\
\phi' &=&  - N \, \partial_{\phi} \lambda_{2}, \label{c2:
BPS-matching2-s2}
\end{eqnarray}
at the second brane $\Sigma_{2}$. In the previous expressions we have defined the
4-D energy-momentum tensors $T^{1}_{\mu \nu}$ and $T^{2}_{\mu \nu}$ describing
$\Psi_{1}$ and $\Psi_{2}$ in the conventional way
\begin{eqnarray}
T^{1}_{\mu \nu} = - \frac{2}{\sqrt{-g}}\frac{\delta S_{1}}{\delta
g^{\mu \nu}} \bigg|_{z_{1}} \qquad \mathrm{and} \qquad T^{2}_{\mu
\nu} = - \frac{2}{\sqrt{-g}}\frac{\delta S_{2}}{\delta g^{\mu
\nu}} \bigg|_{z_{2}}, \label{c2: def-t-munu}
\end{eqnarray}
where $S_{1}$ and $S_{2}$ are the terms appearing in the action
$S_{\mathrm{brane}}$ of Eq. (\ref{c2: brane-matter}).

\subsection{BPS solutions} \label{sec: BPS sol}

Let us, for a moment, assume that the supersymmetry breaking potentials 
$v_{1}(\phi^{1})$ and $v_{2}(\phi^{2})$ defined in Section
\ref{supergravity} and brane matter fields $\Psi_{1}$ and $\Psi_{2}$  
are absent. Then, given a superpotential $W(\phi)$,
the bulk scalar field potential and brane tensions become
\begin{eqnarray}
U= (\partial_{\phi} W)^{2} - W^{2},
\qquad \lambda_{1} = W (\phi^{1}), \qquad \mathrm{and} \qquad
\lambda_{2} = - W (\phi^{2}). \label{BPS 2}
\end{eqnarray}
Under these conditions the system presents an important property 
which shall be exploited heavily during the rest of the paper: The system has 
a BPS vacuum state consisting of a static bulk background 
in which branes can be allocated anywhere, without obstruction.
Indeed, suppose that the bulk fields depend only on $z$, and write 
$g_{\mu \nu} = \omega^{2}(z) \eta_{\mu \nu}$, where $\eta_{\mu \nu}$
is the Minkowski metric, then one finds that the entire system of 
equations (\ref{c2: Grav-eq1})-(\ref{c2: scalar}) are solved by 
functions $\omega(z)$ and $\phi(z)$ satisfying
\begin{eqnarray}
\frac{\omega'}{\omega} = -\frac{N}{4} W \qquad \mathrm{and} 
\qquad \phi' = N \partial_{\phi} W . \label{BPS-equ}
\end{eqnarray} 
Remarkably, boundary conditions 
(\ref{c2: Israel-matching2-s1})-(\ref{c2: BPS-matching2-s2})
are also given by these two equations. Thus, the presence of the branes
forces the system to aquire a domain-wall-like vacuum background, 
instead of a flat 5-D Minkowski background. This property allows us to handle 
the complicated system of equations (\ref{c2: Grav-eq1})-(\ref{c2: scalar})
by linearizing fields about this state. This will be considered in detail in
Section \ref{linear}.  

Notice that the warp factor $\omega(z)$ may be solved and
expressed as a function of $\phi(z)$ instead of $z$
\begin{eqnarray}
\omega(\phi) = \exp \left[ -\frac{1}{4}
\int^{\phi} \!\!\! \alpha^{\! -1}(\phi) \, d\phi
\right], \qquad \mathrm{where} 
\qquad \alpha(\phi) \equiv \frac{ \partial_{\phi} W}{W}.
\label{warp}
\end{eqnarray}

\subsubsection{Dilatonic braneworlds} \label{dilatonic}

In the case of dilatonic braneworlds one has $W = \Lambda \, e^{\alpha \phi}$, where 
$\Lambda$ is a mass scale expected to be of order $M_{5}$, and $\alpha$ 
is a dimensionless constant. In this 
case the relations of Eq. (\ref{BPS-equ}) permit us to solve the background 
values $\phi$ and $\omega$
in terms of $z$. Using the gauge $N=1$ for definiteness and assuming $\Lambda>0$, 
one obtains
\begin{eqnarray}
\phi(z) &=& \phi_{1} - \frac{1}{\alpha} \ln \left[1 - \alpha^{2} W_{0} z \right], \\
\omega(z) &=& \left[ 1 - \alpha^{2} W_{0} z \right]^{1/4\alpha^{2}},
\end{eqnarray}
where $\phi_{1} \equiv \phi(0)$ and $W_{0} = \Lambda \, e^{\alpha \phi_{1}}$. Notice 
the presence of a singularity $\omega=0$ at $z = 1/ \alpha^{2} W_{0}$.
Without loss of generality, one may take the position of the first brane 
$\Sigma_{1}$ at $z=0$ (since $\Lambda>0$, this is a positive tension brane). Then, 
the second brane $\Sigma_{2}$ can be anywhere between $z=0$ and 
$z = 1/ \alpha^{2} W_{0}$. Later on, we shall study the case in which $\Sigma_{2}$
is stabilized at the singularity.

\subsection{Effective theory} \label{effective}

To finish, we present the effective theory describing the
dynamics for the zeroth-order fields from the 4-D point of view. The effective theory is
a bi-scalar tensor theory of gravity of the form \cite{Palma:2004et, Palma:2004fh}
\begin{eqnarray}
S &=& \frac{1}{4 \pi G_{*}} \int d^{4}x \sqrt{- g} \bigg[
\frac{1}{4} R - \frac{1}{2} g^{\mu \nu} \gamma^{a b}
\partial_{\mu} \omega_{a} \partial_{\nu} \omega_{b} - V \bigg]  \nonumber\\
&& + S_{1}[\Psi_{1}, A_{1}^{2}  g_{\mu \nu}] + S_{2}[\Psi_{2},
A_{2}^{2}  g_{\mu \nu}], \qquad \label{c3: EFF Action2}
\end{eqnarray}
where $\omega_{a}$, with $a=1,2$, are the values of the warp factor 
$\omega(z)$ at the brane positions $z_a$ and $G_{*}^{-1} \equiv 16 \pi M_{5}^{2}$. 
Observe that $\omega_{a}$ can be expressed in terms of $\phi^{a}$
(the boundary values of $\phi$) by using Eq. (\ref{warp}) evaluated at
$\phi = \phi^{a}$. The elements of the sigma model metric $\gamma^{a b}$ are given by
\begin{eqnarray}
\gamma^{1 1} = - \frac{6 M_{5} }{B^{2} W_{1}} \bigg[ 1 - \frac{2 M_{5}
A_{1}^{2}}{W_{1}} \bigg] ,  \qquad  \gamma^{2 2} = + \frac{6 M_{5}
}{B^{2} W_{2}}
\bigg[ 1 + \frac{2 M_{5} A_{2}^{2}}{W_{2}} \bigg] , \nonumber \\
 \gamma^{1 2} = \gamma^{2 1} = - \frac{12 M_{5}^{2} \omega_{1}
\omega_{2}}{B^{4} W_{1} W_{2}}, \qquad \qquad \qquad \qquad
\end{eqnarray}
where $A_{1}^{2} = \omega_{1}^{2} / B^{2}$, $A_{2}^{2} = \omega_{2}^{2} / B^{2}$, and
\begin{eqnarray}
B^{2} = M_{5} \int_{z_{1}}^{z_{2}} \!\! dz N \omega^{2} = - 4 M_{5} \int_{\omega_{1}}^{\omega_{2}} \!\! d\omega \frac{\omega}{W}.
\end{eqnarray}
Finally, the effective potential $V$ is found to be
\begin{eqnarray}
V(\phi^{1}, \phi^{2}) = 
\frac{3 M_{5}}{8} \left[ A_{2}^{4} v_{2} + A_{2}^{4} v_{1} \right].
\end{eqnarray}
This effective theory can be deduced either by solving the full set of
Eqs. (\ref{c2: Grav-eq1})-(\ref{c2: scalar}) at the linear level
\cite{Palma:2004fh}, or by
directly integrating the extra-dimensional coordinate $z$ in the action
(\ref{c2: tot-act}) using the moduli-space-approximation approach \cite{Palma:2004et}.
We should mention here that Newton's constant, as measured 
by Cavendish experiments on the positive tension brane, 
is given by $G_{\mathrm{N}} = G_{*} A_{1}^{2} = G_{*} \omega_{1}^{2} / B^{2} $.

\section{Linearized gravity} \label{linear}

In this section we deduce the equations of motion governing
the low energy regime of the system --close to the BPS configuration 
presented in Section \ref{sec: BPS sol}-- 
and consider the problem of stabilizing the 
moduli. Our approach will be to linearize gravity by defining a set of
perturbation fields about the aforementioned static vacuum configuration. 

\subsection{Low energy regime equations}

To start with, assume the existence of background fields $\phi_{0}$,
$\omega_{0}$ and $N_{0}$, depending on both $x^{\mu}$ and $z$, and
satisfying the following equations
\begin{eqnarray}
\frac{\omega_{0}'}{\omega_{0}} = 
- \frac{1}{4} N_{0} W(\phi_{0}), \qquad \mathrm{and} \qquad
\phi_{0}'  =  N_{0} \partial_{\phi} W_{0}.
\end{eqnarray}
The bulk scalar field boundary values are
defined to satisfy $\phi_{0}^{1}(x) = \phi_{0}(x,z_{1})$ and
$\phi_{0}^{2}(x) = \phi_{0}(x,z_{2})$. The form of the warp factor
$\omega_{0}$ is already known to us
\begin{eqnarray}
\omega_{0} (z,x) = \exp \left[ -\frac{1}{4}
\int_{\phi_{0}^{*}}^{\phi_{0}} \alpha^{\! -1}(\phi) \, d\phi
\right],
\end{eqnarray}
where $\phi_{0}^{*}$ is an arbitrary constant. Now, we
would like to study the system perturbed about the BPS configuration
of Section \ref{sec: BPS sol}. To this extent, we define the following set of
variables $h_{\mu \nu}$, $\varphi$ and $\psi$, as
\begin{eqnarray}
g_{\mu \nu} &=& \omega_{0}^{2} \tilde g_{\mu \nu}
+ h_{\mu \nu}, \\
\phi &=& \phi_{0} + \varphi, \\
N &=& N_{0} \, e^{\psi},
\end{eqnarray}
where $g_{\mu \nu}$, $\phi$ and $N$ satisfy the equations of motion
(\ref{c2: Grav-eq1}), (\ref{c2: Grav-eq3}) and (\ref{c2: scalar}),
taking into account the presence of matter in the branes and the small
supersymmetry breaking potentials $v_{1}$ and $v_{2}$. Additionally, $\tilde g_{\mu
\nu}$ is defined to depend only on the spacetime coordinate $x$.
The functions $h_{\mu \nu}$, $\varphi$
and $\psi$ are linear perturbations satisfying $|h_{\mu \nu}| \ll \omega_{0}^{2} |
\tilde g_{\mu \nu}|$, $|\varphi| \ll |\phi_{0}|$ and $|\psi| \ll 1$.
Now, if we insert these definitions back into the equations of motion
(\ref{c2: Grav-eq1}), (\ref{c2: Grav-eq3}) and (\ref{c2: scalar}),
and neglect second order quantities in $h_{\mu \nu}$, $\varphi$ and
$\psi$ we obtain the required equations of motion for the low energy
regime: First, Eq. (\ref{c2: Grav-eq1}) leads to
\begin{eqnarray}
W(\phi_{0}) \left[ h'
+ \frac{N_{0} W_{0}}{2} h \right] + 2
\omega_{0}^{2} \frac{\partial W}{\partial \phi_{0}} \varphi' -
N_{0} \omega_{0}^{2} \left[ 2 U_{0} \psi + \frac{\partial
U_{0}}{\partial \phi_{0}} \varphi \right] = N_{0} (X_{0} + \bar X).
\label{c3: lin-A}
\end{eqnarray}
Equation (\ref{c2: Grav-eq3}) leads to
\begin{eqnarray}
h_{\mu \nu}'' - \tilde g_{\mu \nu}  h''  
-  \frac{\partial N_{0}}{\partial \phi_{0}} \frac{\partial
W}{\partial \phi_{0}}  (h_{\mu \nu}' - \tilde g_{\mu \nu}
h') 
+\frac{N_{0}^{2}}{4} \left[ 2 (\partial_{\phi} W_{0} )^{2} -
W_{0}^{2} \right] (h_{\mu \nu} - \tilde
g_{\mu \nu} h) \nonumber\\
- \frac{3}{2} N_{0} \omega_{0}^{2} \left[ W_{0} \, \psi' +
\frac{\partial W}{\partial \phi_{0}} \, \varphi' +  N_{0} U_{0}
\psi + \frac{N_{0}}{2} \frac{\partial U_{0}}{\partial \phi_{0}}
\varphi \right] \tilde g_{\mu \nu} = 2 N_{0}^{2} 
(Y_{\mu \nu}^{0} + \bar Y_{\mu \nu}).
\label{c3: lin-B}
\end{eqnarray}
And finally Eq. (\ref{c2: scalar}) gives
\begin{eqnarray}
\varphi'' - \left[ N_{0} W_{0} + \frac{\partial N_{0}}{\partial
\phi_{0}} \frac{\partial W}{\partial \phi_{0}} \right] \varphi'
+ \frac{1}{2} \frac{N_{0}}{\omega_{0}^{2}} \frac{\partial
W}{\partial \phi_{0}} \left[ h' +
\frac{N_{0} W_{0}}{2} h \right]  \nonumber\\ - N_{0}
\frac{\partial W}{\partial \phi_{0}} \psi' -
\frac{N_{0}^{2}}{2} \left[ 2  \frac{\partial U_{0}}{\partial
\phi_{0}} \psi + \frac{\partial^{2} U_{0}}{\partial \phi_{0}^{2}}
\varphi \right]  = \frac{N^{2}_{0}}{\omega_{0}^{2}} (Z_{0} + \bar Z ). \label{c3: lin-C}
\end{eqnarray}
In the previous equations $U_{0} = U(\phi_{0})$ and $W_{0} =
W(\phi_{0})$. Notice that the trace $h = \tilde g^{\mu \nu} h_{\mu \nu}$ 
is taken with respect to $\tilde g_{\mu \nu}$ instead of $g_{\mu \nu}$. 
Equations (\ref{c3: lin-A}) and (\ref{c3: lin-B}) correspond to the linearized
5-D Einstein equations, while Eq. (\ref{c3: lin-C}) corresponds to
the linearized bulk scalar field equation. 
Notice the appearance of the sums $X_{0} + \bar X$, $Y_{\mu \nu}^{0} + \bar Y_{\mu \nu}$ 
and $Z_{0} + \bar Z$ at the right hand side of Eqs. (\ref{c3: lin-A})-(\ref{c3: lin-C}).  
The quantities $X_{0}$, $Y_{\mu \nu}^{0}$ and $Z_{0}$ are
\begin{eqnarray}
 X_{0} &=& (\tilde \nabla \phi_{0})^{2} - \frac{4}{3}
\tilde R + 8 \,
\omega_{0}^{-1} \tilde \Box \omega_{0},  \\
 Y^{0}_{\mu \nu} &=& \tilde G_{\mu \nu} + \frac{3}{4} \bigg[
\frac{1}{2} \tilde g_{\mu \nu} (\tilde \nabla \phi_{0})^{2} -
\partial_{\mu} \phi_{0} \partial_{\nu} \phi_{0} \bigg]
 + (N_{0} \omega_{0}^{2})^{-1}
\Big[ \tilde g_{\mu \nu} \tilde \Box (N_{0} \omega_{0}^{2})
\nonumber\\ &&  - \tilde \nabla_{\mu} \tilde \nabla_{\nu} (N_{0}
\omega_{0}^{2})  - 3 \tilde g_{\mu \nu}
\partial^{\alpha} \omega_{0}
\partial_{\alpha} (N_{0} \omega_{0}) - 3 \partial_{\mu} \omega_{0}
\partial_{\nu} (N_{0} \omega_{0}) \nonumber\\ &&  - 3 \partial_{\nu} \omega_{0}
\partial_{\mu} (N_{0} \omega_{0}) \Big] , \label{c2: Y-0} \\
Z_{0} &=& - \frac{1}{N_{0}} \tilde g^{\mu \nu} \tilde \nabla_{\mu}
(N_{0}
\partial_{\nu} \phi_{0})  - 2 \omega_{0}^{-1}
\tilde g^{\rho \lambda}
\partial_{\lambda} \omega_{0} \partial_{\rho} \phi_{0}  ,
\end{eqnarray}
whereas $\bar X$, $\bar Y_{\mu \nu}$ and
$\bar Z$ are
\begin{eqnarray}
\bar X  &=& \omega_{0}^{-2}  \frac{4}{3} ( \tilde \Box  h  - \tilde \nabla^{\alpha}
\tilde \nabla^{\beta} h_{\alpha \beta}) 
\label{c3: bar-X} , \\
\bar Y_{\mu \nu} &=& \omega_{0}^{-2} \frac{1}{2} \big( \tilde g_{\mu \nu}  \tilde \Box h -
\tilde \nabla_{\mu} \tilde \nabla_{\nu}  h - \tilde \Box h_{\mu \nu} 
+ \tilde \nabla^{\sigma}
\tilde \nabla_{\nu} h_{\sigma \mu} + \tilde \nabla^{\sigma} \tilde \nabla_{\mu} h_{\sigma
\nu} \nonumber\\ && - \tilde g_{\mu \nu} \tilde \nabla^{\alpha} \tilde \nabla^{\beta}
h_{\alpha \beta} \big)  + \tilde g_{\mu \nu} \tilde \Box \psi 
- \tilde \nabla_{\mu}
\tilde \nabla_{\nu} \psi  , \\
\bar Z &=& - \tilde \Box \varphi . \label{c3: bar-Z}
\end{eqnarray}
Operators such as $\tilde \nabla$ and $\tilde \Box$ are constructed out 
of $\tilde g_{\mu \nu}$ instead of $g_{\mu \nu}$. 
In writing $\bar X$, $\bar Y_{\mu \nu}$ and $\bar Z$, we have neglected 
terms involving products between background fields spacetime derivatives, 
such as $\tilde \nabla_{\mu} \omega_{0}$, and perturbation fields spacetime 
derivatives, such as $\tilde \nabla_{\mu} h$. This is justified as we
shall later consider the stabilization of the background fields.
Boundary conditions (\ref{c2: Israel-matching2-s1})-(\ref{c2:
BPS-matching2-s2}) can also be expressed in terms of linear fields. 
At the brane $\Sigma_{a}$, with $a=1,2$, they take the form
\begin{eqnarray}
h_{\mu \nu}' - \tilde g_{\mu \nu}  h' 
+  \frac{N_{0} W}{2} \left[ h_{\mu \nu} - \tilde g_{\mu
\nu} h \right]
= \frac{3}{2} N_{0} \omega_{0}^{2} \left[ W(\phi_{0}) \, \psi +
\frac{\partial W}{\partial \phi_{0}} \, \varphi \right] \tilde
g_{\mu \nu} \nonumber\\  + \frac{3}{2} N_{0} \omega_{0}^{2} v_{a} \tilde g_{\mu
\nu} + \frac{3}{2} N_{0} \omega_{0}^{2} v_{a} \tilde g_{\mu
\nu} \psi+ \frac{3}{2} N_{0} \omega_{0}^{2} \frac{\partial v_{a}}{\partial \phi} \tilde g_{\mu
\nu} \varphi \mp M_{5}^{-3} N_{0} T_{\mu \nu}^{\,a} , \label{c3: lin-bound 1}
\end{eqnarray}
and
\begin{eqnarray}
\varphi' = N_{0} \frac{\partial W}{\partial \phi_{0}} \psi +
N_{0} \frac{\partial^{2} W}{\partial \phi_{0}^{2}} \varphi +
N_{0} \frac{ \partial v_{a}}{\partial \phi_{0}} +
N_{0} \frac{ \partial v_{a}}{\partial \phi_{0}} \psi + N_{0} \frac{
\partial^{2} v_{a}}{\partial \phi_{0}^{2}} \varphi, \label{c3: lin-bound 2}
\end{eqnarray}
where signs $\mp$ stand for the first and second brane respectively.  
Background quantities like $\phi_{0}$ and $N_{0}$ must be
evaluated at $z = z_{a}$ according to the brane. It is also useful to 
recast Eq. (\ref{c2: Grav-eq2}) in terms of linear variables
\begin{eqnarray}
\frac{1}{2 N_{0}} \left[\frac{1}{\omega_{0}^{2}} (\tilde \nabla^{\nu} h_{\mu \nu} - 
\tilde \nabla_{\mu} h) \right]' = \frac{3}{4} (W_{0} \tilde \nabla_{\mu} \psi 
+ \frac{\partial W}{\partial \phi_{0}}  \tilde \nabla_{\mu} \varphi). \label{master}
\end{eqnarray}

\subsection{Gauge freedom} \label{c3 gauge}

It is important to keep in mind that Eqs. (\ref{c2: Grav-eq1})-(\ref{c2: scalar})
were written in a gauge $N^{\mu}=0$. This gauge is appropriate for studying parallel
branes as we are in the present case. In general, given a set of small
arbitrary parameters $\varepsilon_{z}(x,z)$ and
$\varepsilon_{\mu}(x,z)$, it can be shown that the perturbed theory
is invariant under the following set of gauge transformations
\begin{eqnarray}
\psi &\rightarrow& \psi + \frac{1}{N_{0}^{2}} \left[
\varepsilon_{z}' -
\frac{N_{0}'}{N_{0}} \varepsilon_{z} + N_{0} (\nabla^{\mu} N_{0}) \varepsilon_{\mu} \right], \\
N_{\mu} &\rightarrow& N_{\mu} + \partial_{\mu} \varepsilon_{z} +
\varepsilon_{\mu}' - 2 N_{0} K^{\nu}_{\mu} \varepsilon_{\nu} - 2
\frac{\partial_{\mu} N_{0}}{N_{0}} \varepsilon_{z}, \\
h_{\mu \nu} &\rightarrow& h_{\mu \nu} + \nabla_{\mu}
\varepsilon_{\nu} + \nabla_{\nu} \varepsilon_{\mu} + \frac{2}{N_{0}}
K_{\mu \nu} \varepsilon_{z}.
\end{eqnarray}
The gauge parameter $\varepsilon_{\mu}(x,z)$ can be used
to eliminate $N_{\mu}$ from the perturbed theory as we have done.  
The gauge parameter $\varepsilon_{z}(x,z)$ can be
used similarly to redefine (or eliminate) $\psi$. Observe that 
there is a residual gauge freedom to choose $\varepsilon_{\mu}(x,z)$
without spoiling condition $N_{\mu} = 0$. Indeed, if $\varepsilon_{\mu}(x,z)$
satisfies
\begin{eqnarray}
\varepsilon_{\mu}' = 2 N_{0} K^{\nu}_{\mu} \varepsilon_{\nu}, \label{last gauge}
\end{eqnarray}
then we can redefine $h_{\mu \nu}$ and continue keeping $N_{\mu} = 0$. This gauge freedom 
makes zero mode gravity invariant under diffeomorphisms, as it should be.

\subsection{Homogeneous solutions}

Observe that the most general set of solutions
$h_{\mu \nu}$, $\psi$ and $\varphi$ can be written in the following form
\begin{eqnarray}
h_{\mu \nu} = \hat h_{\mu \nu} + \bar h_{\mu \nu}, \qquad 
\psi = \hat \psi + \bar \psi, \qquad \mathrm{and} \qquad  
\varphi = \hat \varphi + \bar \varphi.
\end{eqnarray}
Here, fields $\hat h_{\mu \nu}$, $\hat \psi$ and $\hat \varphi$ are the specific
solutions to the system, i.e. those solutions to Eqs. (\ref{c3: lin-A})-(\ref{c3: lin-C}) 
and boundary conditions (\ref{c3: lin-bound 1})-(\ref{c3: lin-bound 2}) including
the inhomogeneous terms $X_{0}$, $Y_{\mu \nu}^{0}$ and $Z_{0}$. On the other hand,
fields $\bar h_{\mu \nu}$, $\bar \psi$ and $\bar \varphi$ are homogeneous solutions
satisfying  Eqs. (\ref{c3: lin-A})-(\ref{c3: lin-C}) but only with 
$\bar X$, $\bar Y_{\mu \nu}$ and $\bar Z$ at the 
right hand side. They also satisfy the following linear boundary conditions
\begin{eqnarray}
h_{\mu \nu}' - \tilde g_{\mu \nu}  h' +  \frac{N_{0} W}{2} \left[ h_{\mu \nu} - \tilde g_{\mu
\nu}  h \right] 
&=& \frac{3}{2} N_{0} \omega_{0}^{2} \left[ W(\phi_{0}) \, \psi +
\frac{\partial W}{\partial \phi_{0}} \, \varphi \right] \tilde
g_{\mu \nu} \nonumber\\ && + \frac{3}{2} N_{0} \omega_{0}^{2} v_{a} \tilde g_{\mu
\nu} \psi + \frac{3}{2} N_{0} \omega_{0}^{2} \frac{\partial v_{a}}{\partial \phi} \tilde g_{\mu
\nu} \varphi  , \label{c4: lin-bound 1}
\end{eqnarray}
and
\begin{eqnarray}
\varphi' = N_{0} \frac{\partial W}{\partial \phi_{0}} \psi +
N_{0} \frac{\partial^{2} W}{\partial \phi_{0}^{2}} \varphi  +
N_{0} \frac{ \partial v_{a}}{\partial \phi_{0}} \psi + N_{0} \frac{
\partial^{2} v_{a}}{\partial \phi_{0}^{2}} \varphi, \label{c4: lin-bound 2}
\end{eqnarray}
at both branes $a = 1,2$ respectively. Observe that 
matter fields do not appear in this set of boundary conditions.  
It was shown in \cite{Palma:2004fh} that 
the specific solutions $\hat h_{\mu \nu}$, $\hat \psi$ and $\hat \varphi$ 
are related to the zeroth-order fields in a special way: They are generated 
by the evolution of the zeroth-order fields 
$\omega_{0}(x,z)$, $N_{0}(x,z)$ and $\phi_{0}(x,z)$ on
the bulk and branes and, when integrated, they give rise to the effective theory 
shown in Section \ref{effective}. In this article we are utterly interested on
the homogeneous solutions $\bar h_{\mu \nu}$, $\bar \psi$ and $\bar
\varphi$. They appear linearly
coupled to the matter energy momentum tensor $T_{\mu \nu}$ in the brane, 
which is just what we need to compute 
corrections to the Newtonian potential at short distances (see Section \ref{Newtonian pot}).

\subsection{Stabilization of the moduli} \label{stabilization}

It is clear from the effective theory shown in Section \ref{effective} that, 
in the absence of supersymmetry breaking potentials $v_{1}$ and $v_{2}$, 
the scalar fields $\phi_{1}$ and $\phi_{2}$ are massless. 
Recall that $\phi_{1}$ and $\phi_{2}$ are
the boundary values of the bulk field $\phi$ at the branes (we could have equally 
chosen a combination between the radion and only one of the boundary values, 
say $\phi_{1}$). Solar system tests of gravity provide strong 
constraints on the conformal couplings $A_{1}$ and $A_{2}$ 
between the moduli and matter fields (recall Section \ref{effective}),
at the extent of making difficult to reconcile natural values for the parameters of
the model and observations \cite{Palma:2005wm}. For example, in the case of a dilatonic 
superpotential $W(\phi) \propto e^{\alpha \phi}$, solar system tests require
$\alpha^{2} < 1.5 \times 10^{-6}$, 
whereas in 5-D Heterotic M-theory one expects $\alpha^{2} = 3/2$.
For this reason, we consider the stabilization of the moduli by introducing 
boundary supersymmetry breaking terms 
$v_{1}$ and $v_{2}$, implying a potential
\begin{eqnarray}
V(\phi^{1}, \phi^{2}) =  \frac{3 k}{8} \left[ A_{2}^{4} v_{2} + A_{2}^{4} v_{1} \right].
\end{eqnarray}
To be consistent with low energy phenomenology, we shall further 
assume that the moduli are driven by this potential to fixed points such that 
$v_{1}(\phi^{1}) = v_{2}(\phi^{2}) = \partial_{\phi} v_{1}(\phi^{1}) 
= \partial_{\phi} v_{2}(\phi^{2}) = 0$, implying a zero effective cosmological 
constant.\footnote{The pair of conditions $v_{1}(\phi^{1}) =0$ 
and $v_{2}(\phi^{2})=0$ are not strictly necessary, as 
present cosmological observations indicate the
existence of a non-negligible dark energy term in our universe.} 
Under these conditions, the zero mode fields $\phi^{1}$ and $\phi^{2}$ acquire masses 
proportional to $\partial_{\phi}^{2} v_{1}(\phi^{1})$ and 
$\partial_{\phi}^{2} v_{2}(\phi^{2})$, respectively. On the other hand,
the small field $\varphi$ appears coupled to the branes also through terms 
proportional to $\partial_{\phi}^{2} v_{1}(\phi^{1})$ and 
$\partial_{\phi}^{2} v_{2}(\phi^{2})$. As we shall see in the following, 
the presence of these couplings drives the system to a stable configuration
in which scalar perturbation fields satisfy $\varphi = h = 0$, and only the 
traceless and divergence-free part of $h_{\mu \nu}$ is free to propagate in the bulk.

To show this, let us start by fixing the gauge $\psi$ as 
\begin{eqnarray}
\psi = -\alpha \varphi , \label{c4: gauge}
\end{eqnarray}
and define the traceless graviton field 
$\gamma_{\mu \nu} = h_{\mu \nu} - \frac{1}{4} \tilde g_{\mu \nu} h$.
With these considerations in mind, the homogeneous equations of motion become
\begin{eqnarray}
W  \left[ h' + \frac{N W}{2} h \right] + 2
\omega^{2} \frac{\partial W}{\partial \phi} \varphi' -
N \omega^{2} \left[ \frac{\partial U}{\partial \phi} - 2 \alpha U \right] \varphi  
= \frac{N}{\omega^{2}}  ( \tilde \Box  h  - \frac{4}{3} \tilde \nabla^{\alpha}
\tilde \nabla^{\beta} \gamma_{\alpha \beta}) ,  
\label{c4: lin-A} \\
h'' -  \frac{\partial N}{\partial \phi} \frac{\partial W}{\partial \phi}  h' 
+ \frac{N^{2}}{4} \left[ 2 (\partial_{\phi} W)^{2} -
W^{2} \right] h  =  - \frac{1}{2} \frac{N^{2}}{\omega^{2}} 
\big( \tilde \Box h - \frac{4}{3} \tilde \nabla^{\alpha}
\tilde \nabla^{\beta} \gamma_{\alpha \beta}  \big) \nonumber\\ + 2 \alpha N^{2} \tilde \Box \varphi , 
\label{c4: lin-B1} \\
\gamma_{\mu \nu}'' -  \frac{\partial N}{\partial \phi} \frac{\partial
W}{\partial \phi}  \gamma_{\mu \nu}'
+\frac{N^{2}}{4} \left[ 2 (\partial_{\phi} W)^{2} -
W^{2} \right]  \gamma_{\mu \nu}  =  
\frac{N^{2}}{\omega^{2}} \big( \frac{1}{8} \tilde g_{\mu \nu}  \tilde \Box h -
\frac{1}{2} \tilde \nabla_{\mu} \tilde \nabla_{\nu}  h 
\nonumber\\ - \tilde \Box \gamma_{\mu \nu}  
+ \tilde \nabla^{\sigma} \tilde \nabla_{\nu} \gamma_{\sigma \mu} 
+ \tilde \nabla^{\sigma} \tilde \nabla_{\mu} \gamma_{\sigma \nu} 
- \frac{1}{2} \tilde g_{\mu \nu} \tilde \nabla^{\alpha} \tilde \nabla^{\beta}
\gamma_{\alpha \beta} \big) 
\nonumber\\ - \alpha N^{2} \big( \frac{1}{2} \tilde g_{\mu \nu} \tilde \Box \varphi 
- 2 \tilde \nabla_{\mu}  \tilde \nabla_{\nu} \varphi \big) , 
\label{c4: lin-B2} \\
\varphi'' - \left[ N W + \frac{\partial N}{\partial
\phi} \frac{\partial W}{\partial \phi} \right] \varphi' 
+ \frac{1}{2} \frac{N^{2}}{\omega^{2}} \frac{\partial W}{\partial \phi} \left[ h' 
+ \frac{N W}{2} h \right]
+ N \frac{\partial W}{\partial \phi}  (\alpha \varphi)' \nonumber\\
- \frac{N^{2}}{2} \left[ \frac{\partial^{2} U}{\partial \phi^{2}} - 2 \alpha
\frac{\partial U}{\partial \phi}  \right] \varphi   
=  - \frac{N^{2}}{\omega^{2}} \tilde \Box \varphi . \label{c4: lin-C}
\end{eqnarray}
Additionally, boundary conditions acquire the form
\begin{eqnarray}
\gamma_{\mu \nu}' + \frac{N W}{2} \gamma_{\mu \nu} = 0, \\
h ' + \frac{N W}{2} h = 0, \\
\varphi'/N =  \frac{\partial^{2} W}{\partial \phi^{2}} \varphi
- \alpha \frac{\partial W}{\partial \phi} \varphi  +  \frac{
\partial^{2} v_{a}}{\partial \phi^{2}} \varphi,
\end{eqnarray}
at both branes [recall that we are using 
$v_{1}(\phi^{1}) = v_{2}(\phi^{2}) = \partial_{\phi} v_{1}(\phi^{1}) 
= \partial_{\phi} v_{2}(\phi^{2}) = 0$]. In the present gauge, Eq. (\ref{master}) becomes $
[ (\tilde \nabla^{\sigma} \gamma_{\sigma \mu} - \frac{3}{4} \tilde
\nabla_{\mu}  h)/\omega^{2}]' = 0$. This means 
\begin{eqnarray}
\tilde \nabla^{\sigma} \gamma_{\sigma \mu} - \frac{3}{4} \tilde
\nabla_{\mu}  h = \omega^{2} f_{\mu}(x),
\end{eqnarray} 
where $f_{\mu}(x)$ is some vector field independent of coordinate $z$. Notice that the 
$z$-dependence $\propto \omega^{2}$ of the combination
$\tilde \nabla^{\sigma} \gamma_{\sigma \mu} - \frac{3}{4} \tilde
\nabla_{\mu} h$ is the same one of the zero-mode graviton. This allows
us to absorb $\tilde \nabla^{\sigma} \gamma_{\sigma \mu} - \frac{3}{4} \tilde
\nabla_{\mu} h$ in the definition of the zero mode 
graviton $\omega^{2} \tilde g_{\mu \nu}$, 
and set $f_{\mu}(x) = 0$ everywhere in the bulk and 
branes: We achieve this by exploiting the remaining gauge (\ref{last gauge}). 
Then, by evaluating Eq. (\ref{c4: lin-A}) at the boundaries, one obtains
\begin{eqnarray}
\omega^{2} \alpha W \frac{\partial^{2} v_{a}}{\partial \phi^{2}} \varphi \Big|_{z_{1}}
= \omega^{2} \alpha W \frac{\partial^{2} v_{a}}{\partial \phi^{2}} \varphi \Big|_{z_{2}} 
= 0.
\end{eqnarray} 
Thus, unless
$\frac{\partial^{2} v_{1}}{\partial \phi^{2}}$ and 
$\frac{\partial^{2} v_{2}}{\partial \phi^{2}}$ are zero,
$\varphi$ and $\varphi'$ must be null at the boundaries. 
This forces the perturbation field $\varphi$ to stabilize in the 
entire bulk. Strictly speaking, this argument is only valid for 
$\frac{\partial^{2} v_{a}}{\partial \phi^{2}} \gtrsim 
\big| \frac{\partial^{2} W}{\partial \phi^{2}} \big|$. For small values 
$\frac{\partial^{2} v_{a}}{\partial \phi^{2}} \ll 
\big| \frac{\partial^{2} W}{\partial \phi^{2}} \big|$, one has to take
into account higher order terms
in the expansion of $\phi$, $N$ and $g_{\mu \nu}$, and the first order 
perturbation
$\varphi$ would not be stabilized at scales of phenomenological interest. 
This is also true for the case in which the squared mass $m_{\varphi}^{2}$ 
of the $\varphi$ excitation 
is larger than $m_{\phi}^{2} \simeq W_{0} \frac{\partial^{2} v_{a}}{\partial \phi^{2}}$.
With $\varphi$ stabilized, it is easy to check that the graviton trace
$h$ and divergence $\tilde \nabla^{\mu} \gamma_{\mu \nu}$ are also 
reduced to zero.
The only mode not affected by the boundary stabilizing terms 
$v_{1}$ and $v_{2}$ is the 
traceless and divergence-free tensor $\gamma_{\mu \nu}$, which is left satisfying the 
following equation of motion
\begin{eqnarray}
\gamma_{\mu \nu}'' - \frac{\partial N}{\partial \phi} \frac{\partial
W}{\partial \phi} \gamma_{\mu \nu}' + \frac{1}{4}N^{2}[2
(\partial_{\phi} W)^{2} - W^{2}] \gamma_{\mu \nu} &=& -
\frac{N^{2}}{\omega^{2}} \tilde \Box \gamma_{\mu \nu}  , \label{c4:
short3-2}
\end{eqnarray}
and boundary conditions
\begin{eqnarray}
\gamma_{\mu \nu}' + \frac{N W}{2} \gamma_{\mu \nu} = 0.
\label{c4: new bound}
\end{eqnarray}
In the next section we consider solving this equation and show 
how $\gamma_{\mu \nu}$ introduces modifications 
to general relativity at short distances.

\section{Newtonian potential} \label{Newtonian pot}

Now we consider the computation of the Newtonian potential for 
single brane models, in which the second brane $\Sigma_{2}$ is assumed
to be stabilized at the bulk singularity $\omega=0$.
The tensor $\gamma_{\mu \nu}$ described by Eq. (\ref{c4: short3-2}) has 
five independent degrees of freedom. From the four dimensional point of view, 
these are just the necessary degrees of freedom to describe massive gravity.
In fact, as we shall see in the following, there is an infinite tower of massive 
gravitons with masses determined by the boundary conditions at the orbifold 
fixed points. 
First, notice that Eq. (\ref{c4: short3-2}) can be
further simplified: Assume a 4-D Minkowski background, and let 
$\gamma_{\mu \nu} = e^{i p \cdot x} \omega^{1/2} \Phi_{m}(z)$ 
with $p^{2} = - m^{2}$; consider also the gauge $N = \omega$. 
Here $\Phi_{m}(z)$ 
is the amplitude of a Fourier mode representing a graviton state of mass $m$ 
(for simplicity, we are disregarding tensorial indexes). 
This leads us to
the following second order differential equation
\begin{eqnarray}
\left[ - \partial_{z}^{2} + v(z) \right] \Phi_{m} = m^{2} \Phi_{m} ,
\label{c2: spect2}
\end{eqnarray}
where $v(z)$ is given by
\begin{eqnarray}
v(z) = \frac{3}{8} \omega^{2} W^{2} \left[ \frac{5}{8} -
\alpha^{2} \right].
\end{eqnarray}
The boundary conditions for $\Phi_{m}(z)$ are now
\begin{eqnarray}
\Phi_{m}' + \frac{3}{8} \omega W \Phi_{m} = 0, \label{bound again}
\end{eqnarray}
at both branes. Notice that $\Phi_{m}(z)$ defines an orthogonal 
set of fields. Indeed, from Eq. (\ref{c2: spect2}) it directly follows
\begin{eqnarray}
\Phi_{m}(z) \Phi_{n}(z) = \frac{1}{m^{2} - n^{2}} 
\left[ \Phi_{m}(z)  \Phi_{n}''(z) -  \Phi_{m}''(z)  \Phi_{n}(z) \right],
\end{eqnarray}
relation which, after integrating and applying boundary conditions 
(\ref{bound again}), gives
\begin{eqnarray} 
\int_{z_{1}}^{z_{2}} dz \Phi_{m}(z) \Phi_{n}(z) = \delta_{n m},
\end{eqnarray}
provided that the $\Phi_{m}$'s are correctly normalized. 
To solve Eq. (\ref{c2: spect2}) it is necessary to know the precise
forms of $W$ and $\omega$ as functions of $z$. They, of course,
must be solved out from the BPS relations
\begin{eqnarray}
\frac{\omega'}{\omega} = -\frac{1}{4}\omega W, \qquad
\mathrm{and} \qquad \phi' = \omega \frac{\partial W}{\partial
\phi}. \label{c2: spect3}
\end{eqnarray}
The first of these two equations gives
\begin{eqnarray}
\omega^{2} = \left( 1 + \frac{1}{4} \int_{0}^{z} \!\! W dz
\right)^{\!\!-2},
\end{eqnarray}
where we have imposed $\omega(0) = 1$ and 
assumed, without loss of generality, that the first brane is located at $z=0$. Notice that there 
is a rich variety of possibilities for the function $v(z)$, depending
on the form of the superpotential $W(\phi)$. In Section \ref{alpha constant}
we shall focus our efforts on the simple case
of dilatonic braneworlds $W(\phi) = \Lambda \, e^{\alpha \phi}$.
There we find that, depending on the value that $\alpha$
takes, one may have either a continuum spectra of massive gravitons, or a 
discrete tower of states.

\subsection{The potential}

We now compute the short distance effects of bulk gravitons on the Newtonian
potential. For this we consider the ideal case of two point particles 
of masses $m_{1}$ and $m_{2}$ at rest on the
same positive tension brane. To proceed,
observe that the traceless graviton field $\gamma_{\mu \nu}$ comes 
coupled to the brane matter fields through the term
\begin{eqnarray}
\mathcal{L}_{\mathrm{int}} = - \frac{1}{2}  \gamma_{\mu \nu} T^{\mu \nu} \delta(z),
\end{eqnarray}
where $\delta(z)$ is the Dirac delta function about $z=0$. 
It is then possible to show \cite{Callin:2004py} that the Fourier transformation
of the Newtonian potential $V(r)$ describing the interaction between 
two sources with energy momentum tensors  $T_{1}^{\mu \nu}$ and $T_{2}^{\mu \nu}$, 
is given by
\begin{eqnarray}
V(\mathbf{k}) = - \frac{1}{2 M_{5}^{3}} \sum_{m} |\Phi_{m}(0)|^{2} 
\frac{T_{1}^{\mu \nu} P^{(m)}_{\mu \nu \alpha \beta} 
T_{2}^{\alpha \beta} }{ \mathbf{k}^{2} + m^{2}} , \label{Fourier}
\end{eqnarray}
where $\Phi_{m}(0)$ are the normalized graviton amplitudes 
evaluated at $z=0$, and $P^{(m)}_{\mu \nu \alpha \beta}$ is 
the polarization tensor for the graviton mode of mass $m$ \cite{Giudice:1998ck}.  
This result comes from the Kallen-Lehmann spectral representation 
of the graviton propagator. 
In the case of point particles of masses $m_{1}$ and $m_{2}$ at rest one has 
$T_{i}^{\mu \nu}(\mathbf{k}) = m_{i} \delta^{\mu}_{0} \delta^{\nu}_{0}$. This means
that the only relevant components of the polarization tensor are the $0000$ elements
$P^{(0)}_{0000}= 1/2$ for the case of massless gravitons, and $P^{(m)}_{0000}= 2/3$ 
for the case of massive gravitons with $m>0$. Putting all
this together into Eq. (\ref{Fourier}) and Fourier transforming back to coordinate 
space, we obtain
\begin{eqnarray}
V(r) = - \frac{1}{8 \pi M_{5}^{3}} \frac{m_{1} m_{2}}{r} \left[ 
\frac{1}{2} |\Phi_{0}(0)|^{2} + \frac{2}{3} \sum_{m>0} |\Phi_{m}(0)|^{2} e^{-m r}  \right].
\end{eqnarray}
Observe that the zero mode amplitude satisfies 
$|\Phi_{0}|^2 = M_{5} \omega^{2}(0)/B^{2}$, where $B^{2}$ was defined in 
Section \ref{effective}. This gives the right value for the Newtonian constant
$G_{\mathrm{N}} ^{-1}= 16 \pi M^{2}_{5} B^{2}$ as defined in the effective theory 
for the zero mode fields. We thus obtain the general expression
\begin{eqnarray}
V(r) = - G_{\mathrm{N}} \frac{m_{1} m_{2}}{r} \left[ 1 + f(r) \right], 
\end{eqnarray}
where $f(r)$ is the correction to Newton's inverse-square law, defined as
\begin{eqnarray}
f(r)  = \frac{4}{3} |\Phi_{0}(0)|^{-2} \sum_{m>0} |\Phi_{m}(0)|^{2} e^{-m r}. \label{f(r)}
\end{eqnarray}

\subsection{Newtonian potential for $\mathbf{\alpha =}$ constant} \label{alpha constant}

It is possible to compute an exact expression for $f(r)$ 
in the case $\alpha =$constant 
(dilatonic braneworlds). The Randall-Sundrum case is reobtained when $\alpha =0$.
For concreteness, let us take $W(\phi) = \Lambda \, e^{\alpha \phi}$ where 
$\Lambda > 0$ is some fundamental mass scale. Then, 
solutions to Eqs. (\ref{c2: spect3}) are simply
\begin{eqnarray}
\omega (z) &=& \left[ 1 + \frac{1-4\alpha^2}{4} W_{0} z \right]^{-\frac{1}{1-4\alpha^2}}, \\
W (z) &=& W_{0} \left[ 1 + \frac{1-4\alpha^2}{4} 
W_{0} z \right]^{\frac{4 \alpha^{2}}{1-4\alpha^2}}, \label{sol omega-sigma}
\end{eqnarray}
where we have defined $W_{0} = \Lambda \, e^{\alpha \phi_{1}}$, with $\phi_{1}$
the value of $\phi$ at the positive tension brane. The form of $v(z)$ is then remarkably 
simple, depending on the value of $\alpha$. If $\alpha$ is in the range 
$0 < \alpha^{2} < 1/4$, then
\begin{eqnarray}
v(z) = \left[ \nu^{2} - \frac{1}{4} \right] \frac{k^{2}}{(1+kz)^{2}} \qquad 
\mathrm{with} \qquad
\nu \equiv \frac{3}{2} (1-4\alpha^{2})^{-1} + \frac{1}{2},
\end{eqnarray}
where $k \equiv (1 - 4 \alpha^{2}) W_{0}/4$. Notice that, although
the singularity is at a finite proper distance from the positive tension 
brane (as we saw in Section \ref{dilatonic}) in this coordinate system the 
singularity is at $z = + \infty$ (recall that we took $N=\omega$). 
If $\alpha^{2} = 1/4$ then $\omega(z) = e^{-W_{0} z/4}$
and $W(z) = W_{0} e^{W_{0} z/4}$, and the potential $v(z)$ is just
a constant
\begin{eqnarray}
v(z) = \left[ \frac{3 W_{0}}{8} \right]^{2}.
\end{eqnarray}
Observe that in this case the singularity is also at infinity. Finally, if 
$\alpha$ is in the range $1/4 > \alpha^{2}$, then 
\begin{eqnarray}
v(z) = \left[ \nu^{2} - \frac{1}{4} \right] \frac{\mu^{2}}{(1 - \mu z)^{2}} 
\qquad \mathrm{with} \qquad \nu \equiv \frac{3}{2} (4\alpha^{2}-1)^{-1} - \frac{1}{2},
\end{eqnarray}
where $\mu \equiv (4 \alpha^{2} - 1) W_{0}/4$. Observe that in this case
the singularity is at $z = 1/\mu$. In all of these three cases, one has
\begin{eqnarray}
|\Phi_{0}(0)|^{2} =  \frac{1+2\alpha^{2}}{2} W_{0},
\end{eqnarray}
which means that $G_{\mathrm{N}}^{-1} = \frac{32 \pi}{1+2\alpha^{2}} M_{5}^2 W_{0}^{-1}$.
It is interesting to notice that for the cases $0 < \alpha^{2} < 1/4$ and 
$\alpha^{2} = 1/4$ there is a continuum of massive gravitons, irrespective 
of the fact that the extra-dimension is actually finite. This is due 
to the warping of the extra-dimension and the presence of the singularity.
In the following, we find solutions to this system case by case.

\subsubsection{Case $\alpha^{2} = 1/4$}

Let us start with the simplest case. Here $v =  (3 W_{0}/8)^{2}$ is just a constant,
and solutions are given by linear combinations of trigonometric functions. 
The singularity $\omega=0$ is at $z=+\infty$, so it is convenient to keep the 
second brane at a finite position $z=z_{s}$ to impose boundary conditions, 
and then let $z_{s} \rightarrow +\infty$. 
Normalized solutions, satisfying appropriate boundary conditions are 
\begin{eqnarray}
\Phi_{m}(z) = \sqrt{\frac{2}{z_{s} m^{2}}} \left[ \sqrt{m^{2}-v} \cos (\lambda_{m} z) 
+ \sqrt{v} \sin (\lambda_{m} z) \right],
\end{eqnarray}
where $\lambda_{m} = \sqrt{m^{2}-v}$. The masses are quantized as
\begin{eqnarray}
m^{2} = v + \left( \frac{n\pi}{z_{s}} \right)^{2}, \label{n}
\end{eqnarray}
with $n=1,2,3, \cdots$. Equation (\ref{n}) allows to define the appropriate
integration measure over the spectra in the limit  $z_{s} \rightarrow +\infty$.
Indeed, one finds $\sum_{n} \rightarrow \frac{z_{s}}{\pi} \int m/ \sqrt{m^2-v} \, dm$,
with the integration performed between $\sqrt{v}$ and $+\infty$. 
Then, putting it all back together into Eq. (\ref{f(r)}), we obtain
\begin{eqnarray}
f(r) =  \frac{32}{9 \pi W_{0}}  \int_{\sqrt{v}}^{\infty} dm 
\frac{\sqrt{m^{2} - v}}{m} e^{-mr}. \label{f1}
\end{eqnarray}

\subsubsection{Case $\alpha^{2}<1/4$}

Here the singularity is also at $z=+\infty$, so we use the same technique as before,
and let $z_{s} \rightarrow + \infty$ after imposing boundary conditions.
General solutions to (\ref{c2: spect2}) are
\begin{eqnarray}
\Phi_{m} (z) = \sqrt{1 + kz} \left( A_{m} J_{\nu} \left[ \frac{m}{k}(1+kz) \right] 
+ B_{m} Y_{\nu} \left[ \frac{m}{k}(1+kz) \right] \right),
\end{eqnarray}
where $J_{\nu}(x)$ and  $Y_{\nu}(x)$ are the usual Bessel functions of order $\nu$.
Recall that here $\nu = \frac{3}{2} (1-4\alpha^{2})^{-1} + \frac{1}{2}$. 
Boundary conditions
at the first brane position give $A_{m} J_{\nu - 1}  \left[ m/k \right] + 
B_{m} Y_{\nu - 1}  \left[ m/k \right] = 0$, which allows to write 
\begin{eqnarray}
\Phi_{m} (z) = N_{m} \sqrt{1 + kz} \left( J_{\nu - 1}  \left[ m/k \right] 
 Y_{\nu} \left[ \frac{m}{k}(1+kz) \right]  -    Y_{\nu - 1}  \left[ m/k \right] 
J_{\nu} \left[ \frac{m}{k}(1+kz) \right] \right). \nonumber\\ \label{solution}
\end{eqnarray}
Boundary conditions at the second brane $z = z_{s}$ give
\begin{eqnarray}
\frac{Y_{\nu - 1}  \left[ m/k \right]}{J_{\nu - 1}  \left[ m/k \right]} 
= \frac{Y_{\nu - 1}  \left[ (1 + k z_{s} ) m/k \right]}{J_{\nu - 1}  \left[ (1 + k z_{s}) m/k \right]}.
\end{eqnarray}
For $z_{s} \rightarrow +\infty$, this condition implies a quantization of $m$ of 
the form $m_{n} = \frac{n \pi}{z_{s}}$ which, in turn, permits us
to define the integration measure over the spectra as 
$\sum_{m} \rightarrow \frac{z_{s}}{\pi} \int dm$. Now the integration is 
between $0$ and $+\infty$. The normalization constant $N_{m}$ of 
Eq.(\ref{solution}) is found to be
\begin{eqnarray}
N_{m}^{-2} &=&  \int_{0}^{z_{s}} dz (1+kz) 
\left( J_{\nu - 1} [m/k] Y_{\nu} \left[ \frac{m}{k}(1+kz) \right]  
+ Y_{\nu - 1}[m/k] J_{\nu} \left[ \frac{m}{k}(1+kz) \right] \right)^{2} \nonumber\\
&=& \frac{k z_{s}}{\pi m} \left( J_{\nu - 1}^{2} [m/k] + Y_{\nu - 1}^{2}[m/k] \right).
\end{eqnarray}
The second equality comes out in the limit $z_{s}\rightarrow + \infty$.
All of this allows us to compute $f(r)$ by using Eq. (\ref{f(r)})
\begin{eqnarray}
f(r) =  \frac{8}{3\pi^{2}} \frac{1 - 4\alpha^{2}}{1+2\alpha^{2}}  
\int_{0}^{\infty} \frac{dm}{m}   
\frac{e^{-mr}}{ J_{\nu - 1}^{2} [m/k] + Y_{\nu - 1}^{2}[m/k]}, \label{f2}
\end{eqnarray}
where we used the identity 
$J_{\nu - 1}[x] Y_{\nu }[x] - Y_{\nu - 1}[x] J_{\nu }[x] = -2/\pi x$.
It is of interest to check whether  Eq. (\ref{f1}) is reobtained out of 
Eq. (\ref{f2})  by letting $\alpha^{2} \rightarrow 1/4$. This is indeed the
case: It is enough to use the following identity in Eq. (\ref{f2}), 
valid in the limit $\nu \rightarrow +\infty$ (which is equivalent to 
$\alpha^{2} \rightarrow 1/4$)
\begin{eqnarray}
\frac{1}{\nu  \left( J_{\nu - 1}^{2} [x \nu] + Y_{\nu - 1}^{2}[x \nu] \right) }
\rightarrow \frac{\pi}{2} \theta(x-1) \sqrt{x^{2} - 1},
\end{eqnarray}
where $\theta(x-1)$ is the unitary step function about $x=1$.

\subsubsection{Case $\alpha^{2}>1/4 $}

Here the singularity is at a finite coordinate $z=1/\mu$ and
$\nu = \frac{3}{2} (4\alpha^{2}-1)^{-1} - \frac{1}{2}$.
General solutions to Eq. (\ref{c2: spect2}) are of the form
\begin{eqnarray}
\Phi_{m} (z) = N_{m} \sqrt{1 - \mu z} \bigg( 
A_{m} J_{\nu} \bigg[ \frac{m}{\mu}(1-\mu z) \bigg] 
+ B_{m} Y_{\nu} \bigg[ \frac{m}{\mu}(1- \mu z) \bigg] \bigg). 
\end{eqnarray}
Boundary conditions at $z=0$ and $z=1/\mu$ require $B_{m} = 0$ and 
$J_{\nu+1}[m/\mu] = 0$. Let $u^{\nu+1}_{n}$ be the $n$-th zero of
the Bessel function $J_{\nu+1}[x]$. Then, quantized graviton masses are given 
by
\begin{eqnarray}
m_{n} = \mu \, u^{\nu+1}_{n},
\end{eqnarray}
and normalized solutions are easily found to be
\begin{eqnarray}
\Phi_{m}(z) = \sqrt{1-\mu z}  \frac{ \sqrt{2 \mu}}{ J_{\nu+2} [m/\mu]}  
J_{\nu} \bigg[ \frac{m}{\mu}(1-\mu z) \bigg].
\end{eqnarray}
It should be noticed that solutions with $m^{2} < 0$, in principle allowed 
by Eq. (\ref{c2: spect2}) in the range $5/8 < \alpha^{2}$, are discarded 
by boundary conditions. This means that there are no tachionic states in the 
graviton spectra, as it should be.\footnote{It would be interesting 
to investigate whether this result can be extended to any type of 
superpotential $W(\phi)$ involved in
solutions of Eq. (\ref{c2: spect2}) with boundary 
conditions (\ref{bound again}).} Then, $f(r)$ is simply found to be
\begin{eqnarray}
f(r) = \frac{4}{3} \frac{4 \alpha^{2} -1}{1 + 2 \alpha^{2}} 
\sum_{n} e^{- \mu \, u^{\nu+1}_{n} r}. \label{f3}
\end{eqnarray}
Interestingly, this corresponds to a tower of massive states contributing 
Yukawa-like interactions to the Newtonian potential, all of them 
coupled to matter with the same strength.
Again, we should check that Eq. (\ref{f1}) is reobtained out from Eq. (\ref{f3})
in the limit $\alpha^{2} \rightarrow 1/4$. This is possible by noticing that
in the limit $\nu \rightarrow + \infty$, the following relation involving 
Bessel zeros $u^{\nu+1}_{n}$ is satisfied
\begin{eqnarray}
n = \frac{\nu}{\pi} \left[ \sqrt{(u^{\nu}_{n}/\nu)^2-1} 
+ \arctan \left(1/ \sqrt{(u^{\nu}_{n}/\nu)^2-1}\right) \right] - \frac{2 \nu -1}{4}.
\end{eqnarray}
This relation allows us to define the integration measure over the graviton spectra
\begin{eqnarray}
\sum_{n} \longrightarrow 
\frac{4}{\pi W_{0} (4 \alpha^{2}-1)} \int dm \frac{\sqrt{m^{2}-v}}{m},
\end{eqnarray}
which is enough to recover Eq. (\ref{f1}).

\subsection{Short distance corrections}

The only length scale available in the present system, 
apart from the Planck scale, is $W_{0}^{-1}$. Then, we may ask what the 
leading correction to the Newtonian potential is 
in the regime $r W_{0} \gg 1$. Since such correction would be the first 
signature expected from these models at short distance tests of gravity, 
their computation allows us to place phenomenological constraints on 
the values of $W_{0}$.

\subsubsection{Case $\alpha^{2} = 1/4$}

In the case $\alpha^{2} = 1/4$
one finds directly, by using  $r W_{0} \gg 1$ in 
Eq. (\ref{f1})
\begin{eqnarray}
V(r) = - G_{\mathrm{N}} \frac{m_{1} m_{2}}{r} \left[ 1 
+ \frac{8 \pi}{3} \left( \frac{4}{3 \pi W_{0} r} \right)^{3/2} 
\!\! (1 - 3/ W_{0} r) \, e^{-3 r W_{0}/8}   \right]. \label{corr1}
\end{eqnarray}
At present, we know of no constraints on this type of corrections to the
Newtonian potential.

\subsubsection{Case $\alpha^{2}<1/4$}

In the case $\alpha^{2} < 1/4$ one can expand the Bessel functions 
in the small argument limit to find
\begin{eqnarray}
V(r) = - G_{\mathrm{N}} \frac{m_{1} m_{2}}{r} \left[ 1 
+ \frac{8}{3} \frac{1-4\alpha^{2}}{1+2\alpha^{2}} \frac{B[\nu-1,\nu-1]}{(2kr)^{2(\nu-1)}}
\right], \label{corr2}
\end{eqnarray}
where $k = \frac{W_{0}}{4} (1-4 \alpha^{2})$, 
$\nu = \frac{3}{2} (1-4\alpha^{2})^{-1} + \frac{1}{2}$, and 
$B [x,y] = \Gamma[x] \Gamma[y]/ \Gamma[x+y]$ is the usual beta function. Constraints
on $k$ and $\alpha$ for a few values of $2(\nu-1) = 1,2,3, \cdots$ appearing 
in the power law correction $r^{2(\nu-1)}$
can be found in ref. \cite{Adelberger:2006dh}.

\subsubsection{Case $\alpha^{2}>1/4 $}

Finally, in the case $\alpha^{2}>1/4$, one may just pick up the leading contributing 
term involving the first root $u^{\nu+1}_{0}$. This gives a Yukawa force 
correction of the form
\begin{eqnarray}
V(r) = - G_{\mathrm{N}} \frac{m_{1} m_{2}}{r} \left[ 1 
+ \frac{4}{3} \frac{4 \alpha^{2} -1}{1 + 2 \alpha^{2}} e^{- \mu \, u^{\nu+1}_{0} r}
\right]. \label{corr3}
\end{eqnarray}
It is interesting here to consider the particular case of 5-D Heterotic M-theory,
where $\alpha^{2} = 3/2$. In this case, the leading correction to the Newtonian 
potential gives
\begin{eqnarray}
V(r) = - G_{\mathrm{N}} \frac{m_{1} m_{2}}{r} \left[ 1 
+ \frac{5}{3} e^{- r / \lambda}
\right],
\end{eqnarray}
where $\lambda^{-1} \simeq 4.45 W_{0}$.  
Current tests of gravity at short distances \cite{Kapner:2006si} 
provide the constraint $\lambda \le 50\mu$m.

\subsection{Extra-dimensions in the near future?}

A sensible question regarding this type of model is whether there are any chances of
observing short distance modifications of general relativity in the near future. 
To explore this, notice that the relevant energy scale at which corrections
to the conventional Newtonian potential become 
significant is $W_{0} = \Lambda \, e^{\alpha \phi_{1}}$, instead of the 
more fundamental mass scale $\Lambda$. Typically one would expect 
$\Lambda \simeq M_{5}$ which has to be above TeV scales to agree with
particle physics constraints \cite{Mirabelli:1998rt, Hewett:1998sn, Giddings:2001bu, Dimopoulos:2001hw}. 
Nevertheless, the factor
$e^{\alpha \phi_{1}}$ in front of $\Lambda$ leaves open the possibility 
of bringing $\lambda = W_{0}^{-1}$ up to micron scales, depending on the 
vacuum expectation value of $\phi_{1}$. 

In the case of 5-D Heterotic M-theory, for instance, one has
$e^{\alpha \phi_{1}} = 1/\mathcal{V}$, where $\mathcal{V}$ is the 
volume of the Calabi-Yau 3-fold in units of $M_{5}$.
In order to have an accessible scale $\lambda \simeq 10 \mu$m, 
it would be required 
\begin{eqnarray}
\mathcal{V} \, \frac{M_{\mathrm{Pl}}}{M_{5}} \simeq 10^{29}, \label{estim}
\end{eqnarray}
where we assumed $\Lambda \simeq M_{5}$. 
On the other hand, Newton's constant also comes determined by 
a combination of $W_{0}$ and $M_{5}$ in the form 
$G_{\mathrm{N}}^{-1} = \frac{32 \pi}{1+2\alpha^{2}} M_{5}^2 W_{0}^{-1}$. This
implies $M_{\mathrm{Pl}}^{2} \simeq M_{5}^{2} \mathcal{V}$. Thus, to 
achieve the estimation of Eq. (\ref{estim}) one
requires the following values for $M_{5}$ and $\mathcal{V}$
\begin{eqnarray}
M_{5} \simeq 10^{-10} M_{\mathrm{Pl}}, \qquad \mathrm{and} \qquad 
\mathcal{V}^{1/6} \simeq 10^{3}.
\end{eqnarray}
Although, these figures do not arise naturally within string theory, they are in
no conflict with present phenomenological constraints coming from 
high energy physics. In particular, non-zero Kaluza-Klein modes coming from 
the compactified volume $\mathcal{V}$ would have masses of order $10^{6}$GeV.
On the other hand, if $M_{5}$ is of the order of the grand unification scale 
$M_{\mathrm{GUT}} \sim 10^{16}$GeV, then corrections to the Newtonian potential
would be present at the non-accessible scale $\lambda \sim 10^{-20} \mu$m.

\section{Conclusions} \label{Conclusions}

Braneworld models provide a powerful framework to
address many theoretical problems and phenomenological issues, with
a high degree of predictive capacity. They allow a consistent
picture of our four-dimensional world, and yet, they grant the very
appealing possibility of observing new physical phenomena just
beyond currently accessible energies.

In this paper we investigated the gravitational interaction between massive
bodies on the braneworld within a supersymmetric braneworld scenario
characterized by the prominent role of a bulk scalar field.
By doing so, we have learned that it is possible to obtain
short distance modifications to general relativity in ways that differ 
from the well known Randall-Sundrum case. These modifications
represent a distinctive signature for this class of models that 
can be constrained by current tests of gravity at short distances.

The setup considered here consisted of a fairly general class
of supersymmetric braneworld models with a bulk scalar field $\phi$ and
brane tensions proportional to the superpotential $W(\phi)$ of the
theory. For $W(\phi) \neq$ constant, the vacuum state of the 
theory is in general different from the usual AdS profile. Nevertheless, 
in order to have significant effects at short distances 
--say, micron scales-- different from the Randall-Sundrum
case, it was necessary to stabilize the bulk scalar field in a 
way that would not spoil the geometry of the extra-dimensional 
space. To this extent, we considered the inclusion of supersymmetry 
breaking potentials on the branes. After this, the only relevant 
degrees of freedom on the bulk consisted of a massive spectrum of 
gravitons, with masses determined by boundary conditions 
on the branes. 

On the phenomenological side, the main results of this 
article are summarized by Eqs. (\ref{corr1}), (\ref{corr1})
and (\ref{corr3}). They show the leading contributions to 
the Newtonian potential within dilatonic braneworld scenarios, 
which is what would be observed if the tension
of the brane is small enough.
Regarding these results, we indicated that it is plausible 
to expect new phenomena at micron 
scales without necessarily having conflicts with current high energy constraints: 
In the case of dilatonic braneworlds, for example, the necessary value of the 
5-D fundamental scale was $M_{5} \simeq 10^{10} M_{\mathrm{Pl}}$. 
Additionally, in the case of 5-D 
Heterotic M-theory, where $\phi$ is related to the volume of 
small compactified extra-dimensions, we found that Kaluza-Klein modes 
are expected to be of order $\sim 10^{6}$GeV.

\begin{center}

{\bf Acknowledgments}

\end{center}

I would like to thank Fernando Quevedo, Ed Copeland and
Ioannis Papadimitriou for useful comments and discussions.
This work was supported by the Collaborative Research Centre 676
(Sonderforschungsbereich 676) Hamburg, Germany.

\end{document}